# Fracture and fatigue of entangled and unentangled polymer networks

Dongchang Zheng[1†], Shaoting Lin[1†], Jiahua Ni[1], Xuanhe Zhao[1,2*]

[1]Department of Mechanical Engineering, Massachusetts Institute of Technology, Cambridge, MA, USA;
[2]Department of Civil and Environmental Engineering, Massachusetts Institute of Technology, Cambridge, MA, USA

†These authors contributed equally to this work. *Corresponding author: zhaox@mit.edu

**Abstract**

Entanglement of polymer chains is ubiquitous in elastomers, gels, and biological tissues. While the effects of chain entanglement on elasticity and viscoelasticity of polymer networks have been intensively studied, it remains elusive how chain entanglement affects fracture and fatigue of polymer networks. In this paper, using polyacrylamide hydrogels as a model material, we systematically compare fracture toughness and fatigue threshold of polymer networks with various levels of chain entanglement. We find that the fracture toughness and fatigue threshold of an unentangled polymer network are almost the same, although the unentangled polymer network still contains non-ideal features including topological defects (i.e., dangling chains and cyclic loops) and structural heterogeneity (i.e., non-uniform chain lengths and non-uniform functionalities). In contrast, the fracture toughness of an entangled polymer network can be over ten times (up to 16 times) higher than its fatigue threshold, indicating substantial toughness enhancement due to chain entanglement. Different from the conventional toughness enhancement due to bulk dissipation of polymer networks, the toughness enhancement by chain entanglement requires low stress-stretch hysteresis (<10%) of the bulk entangled polymer networks. We attribute the toughness enhancement in entangled polymer networks to a new dissipation mechanism, *near-crack dissipation*, which is possibly induced by pull-out of chains and delocalized damage of chains around the crack tip.

**Keywords**

Fracture, fatigue, chain entanglement, toughening, hysteresis

# 1. Introduction

Entanglement of polymer chains refers to the topological constraints that restrict the molecular motion of neighboring polymer chains [1]. Chain entanglement is ubiquitous in various soft materials including elastomers, gels, and biological tissues. The presence of chain entanglement in polymer networks impacts many mechanical and physical properties of soft materials. For example, chain entanglement can increase rigidity of polymer networks by imposing topological constraints on both crosslinks and polymer chains [2, 3]. Chain entanglement can also induce viscoelasticity of polymer networks due to the reptation of entangled polymer chains [1]. While the effects of chain entanglement on elasticity and viscoelasticity have been intensively studied, its effects on fracture and fatigue of polymer networks have not been well understood.

Fracture and fatigue are two important modes of mechanical failures of soft materials (**Fig. 1**). As a fatigue crack propagates in a soft material under cyclic loads, the measured fatigue threshold $\Gamma_{fatigue}$ accounts for the intrinsic fracture energy, i.e. the energy required to fracture a layer of polymer chains [4]. As a fracture crack propagates in a soft material under a monotonic load, the measured fracture toughness $\Gamma_{fracture}$ accounts for both the intrinsic fracture energy and the energy dissipated in the process zone around the crack tip [4]. The ratio of the fracture toughness to the fatigue threshold of a soft material gives its toughness enhancement $\Gamma_{fracture}/\Gamma_{fatigue}$. The fracture toughness and fatigue threshold have been measured for various soft tough materials, including vulcanized rubbers [5], double-network hydrogels [6, 7], interpenetrating tough hydrogels [8, 9], viscoelastic polyampholyte hydrogels [10, 11], and semi-crystalline hydrogels [12-14]. Their toughness enhancement can be as high as thousands of times. The high toughness enhancement of these soft tough materials typically relies on their large stress-stretch hysteresis [7, 9, 11, 15-17] up to 90% [11]. The stress-stretch hysteresis is defined as the ratio of the dissipated mechanical energy to the total mechanical work done to the material [15, 18]. As a fracture crack propagates in such a soft material under a monotonic load, the large stress-stretch hysteresis in the process zone around the crack dissipates substantial mechanical energy, thereby toughening the material.

Chain entanglement usually gives low stress-stretch hysteresis of bulk polymer networks. For example, the maximum stress-stretch hysteresis reported so far is below 10% for polyacrylamide (PAAm) hydrogels, below 4% for unfilled natural rubber, and below 8% for polydimethylsiloxane (PDMS, Sylgard 184) [19, 20]. Intriguingly, these low-hysteresis soft materials still demonstrate high toughness enhancement. For example, Lake *et al*. [5] and Rivlin *et al*. [21] measured the fatigue threshold and fracture toughness of low-hysteresis unfilled vulcanized natural rubber as 50 J/m$^2$ and 3,700 J/m$^2$, respectively. More recently, Tang *et al*. [22], Zhang *et al*. [14], and Yang *et al*. [19] measured the fatigue threshold and fracture toughness of polyacrylamide (PAAm) hydrogels with various constituents. Despite the low stress-stretch hysteresis ratios of PAAm hydrogels (below 10%), they show high toughness enhancement up to 9.4 [14]. Yang *et al*. further studied the effect of network imperfection in promoting the toughness of PAAm hydrogels; they attributed the high toughness of PAAm hydrogels to non-uniform chain lengths and distributed chain scissions around the crack. Recently, Kim et al. [23] studied the impact of chain entanglement on mechanical properties of polymers. They ascertained that the dense entanglements enable transmission of tension in a polymer chain to many other chains, giving high fracture toughness, high fatigue resistance, low friction, and high wear resistance. Despite these previous works, one important question remains unanswered: how does the low stress-stretch hysteresis of bulk entangled polymer networks give high toughness enhancement, which is contrary to the well-known toughening mechanism for high-hysteresis materials such as ductile metals [24], filled rubbers [25, 26], and tough hydrogels [7, 9]? Notably, previous studies have not systematically tuned the level of chain entanglement in polymer networks, but such systematic tuning of chain entanglement can be critical to answering this question [18].

In this paper, we use PAAm hydrogels as a model material system to investigate the effect of chain entanglement on fracture and fatigue of polymer networks. In order to systematically control the level of chain entanglement, we vary the density of crosslinkers while maintaining the polymer content in the hydrogels. We find that the fracture toughness and fatigue threshold of a nearly unentangled polymer network are almost the same. Since the nearly unentangled polymer network still contains non-ideal features [27] including structural heterogeneity (i.e., non-uniform chain lengths and non-uniform functionalities) and topological defects (i.e., dangling chains and cyclic loops), our experiments reveal that these non-ideal features do not induce significant toughness enhancement of the unentangled polymer network (**Fig. 1a**). We further find that the fracture toughness of an entangled polymer network is 16 times higher than its fatigue threshold, although the maximum stress-stretch hysteresis ratio of the entangled polymer network is lower than 10% (**Fig. 1b**). We attribute the low-hysteresis toughness enhancement in entangled polymer networks to the near-crack dissipation, which is possibly caused by pull-out of chains and delocalized damage of chains around the crack tip. This work not only reveals the effect of chain entanglement on fracture and fatigue of polymer networks but also suggests an effective toughening mechanism for low-hysteresis soft materials.

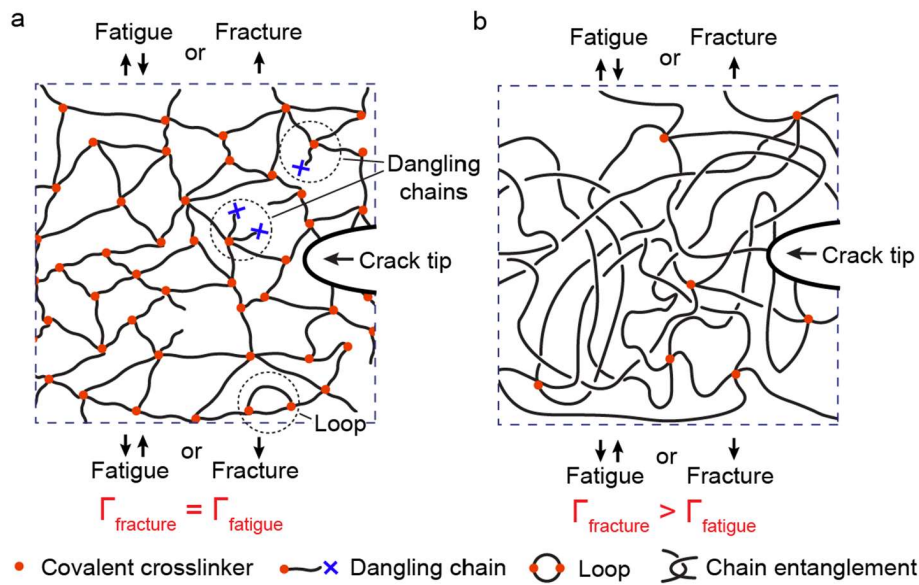

**Fig. 1. Fracture and fatigue of entangled and unentangled polymer networks.** (a) Fracture toughness and fatigue threshold are almost the same for a nearly unentangled polymer network, i.e., $\Gamma_{fracture} = \Gamma_{fatigue}$. While the unentangled polymer network contains non-ideal features including structural heterogeneity (e.g., non-uniform chain lengths, non-uniform functionalities), and topological defects (e.g., dangling chains and cyclic loops), these non-ideal features do not induce significant toughness enhancement of the unentangled polymer network. (b) Fracture toughness is many times larger than fatigue threshold for a highly entangled polymer network, i.e., $\Gamma_{fracture} > \Gamma_{fatigue}$.

## 2. Results and discussions
### 2.1 Fabrication of polymer networks with controlled chain entanglements

We vary the mass fraction of crosslinkers $\phi_c$ (i.e., the ratio of the mass of crosslinkers to the total mass of the hydrogel) while maintaining the mass fraction of polymers $\phi_p$ (i.e., the ratio of the mass of polymers to the total mass of the hydrogel) in the as-prepared state (i.e., reference state) to control the level of chain entanglement in the hydrogels. Given the molecular weight of the monomer as $M_m$ and the molecular weight of the crosslinker as $M_c$, we can calculate the average chain length (i.e., number of monomers) between neighboring crosslinkers in the as-prepared state as $N^{\text{ref}} = (\phi_p M_c)/(\phi_c M_m)$. By further imposing the mass conservation in the hydrogel, we can calculate the average chain density (i.e., number of chains per unit volume of the hydrogel) in the as-prepared state as $n^{\text{ref}} = N_A \rho_g \phi_p / (N^{\text{ref}} M_m)$ with $N_A = 6.02 \times 10^{23}$ mol$^{-1}$ being the Avogadro constant and $\rho_g = 10^3$ kg/m$^3$ being the hydrogel's density. When immersed in a deionized water for a sufficient time, the hydrogel swells and increases its length by a ratio of $\lambda_s$. Notably, the swelling of the hydrogel does not change the average chain length between neighboring crosslinkers, but decreases the average chain density in the hydrogel due to its expanded volume. Hence, the average chain length between neighboring crosslinkers in the hydrogel in the swollen state is the same as that in the reference state, i.e., $N = N^{\text{ref}} = (\phi_p M_c)/(\phi_c M_m)$. The average chain density of the hydrogel in the swollen state can be calculated as $n = n^{\text{ref}} / \lambda_s^3 = (N_A \rho_g \phi_p)/(N^{\text{ref}} M_m \lambda_s^3)$.

We follow the routine protocol for synthesizing polyacrylamide hydrogels via free radical polymerization. 14 g acrylamide monomers (i.e., AAm, A8887, Sigma-Aldrich) with molecule weight of $M_m = 71$ g/mol are first dissolved in 86 ml of deionized water, yielding the solution A. 110 μl 0.1 M Ammonium persulfate (i.e., APS, A3678, Sigma-Aldrich) as the photo initiator, controlled amount (i.e., 165, 300, 600, 1000, 1500, 3000 μl) of 0.23 wt.% N,N'-Methylenebisacrylamide (i.e., Bis-acrylamide, 146072, Sigma-Aldrich) with molecule weight of $M_c = 154$ g/mol as the crosslinker, and 10 μl N,N,N′,N′-Tetramethylethylenediamine (i.e., TEMED, T9281, Sigma-Aldrich) as the accelerator are added into the 13 ml solution A, yielding a final mass fraction of polymers $\phi_p$ as 14 wt.% and controlled mass fraction of crosslinkers $\phi_c$ as 0.0029, 0.0053, 0.0177, 0.0265, 0.0531 wt.%. The pre-gel solution is further filled with nitrogen gas to produce oxygen-free conditions and poured into a rectangular-shaped mold with the dimensions of 40 mm, 20 mm, and 1.5mm. The mold is placed on a hot plate at 50 °C to complete the thermal-induced free radical polymerization. Afterward, the sample is submerged in deionization water to reach its swollen state, measuring its swelling ratio in volume $\lambda_s^3$ ( **Fig. A1**).

At least 24 hours are required to ensure the sample reaching a fully swollen state. The average chain length $N$ and the average chain density $n$ of the hydrogels in the swollen state are summarized in **Fig. 2, a** and **b** for PAAm hydrogels with various amounts of crosslinkers.

Table 1. Summary of notations used in the experiments

| Notation | Definition | Notation | Definition |
|---|---|---|---|
| $\phi_c$ | Ratio of the mass of crosslinkers to the total mass of the hydrogel in the as-prepared state | $\phi_p$ | Ratio of the mass of polymers to the total mass of the hydrogel in the as-prepared state |
| $M_m$ | Molecular weight of the monomer | $M_c$ | Molecular weight of the crosslinker |
| $N^{\text{ref}}$ | Average number of monomers between neighboring crosslinkers in the as-prepared state | $n^{\text{ref}}$ | Average number of chains per unit volume of the hydrogel in the as-prepared state |
| $N_A$ | Avogadro constant | $\rho_g$ | Hydrogel density (taken as water density) |
| $N$ | Average number of monomers between neighboring crosslinkers in the swollen state | $n$ | Average number of chains per unit volume of the hydrogel in the swollen state |
| $\lambda_s$ | Ratio of the hydrogel length in the swollen state to the length in the as-prepared state | | |

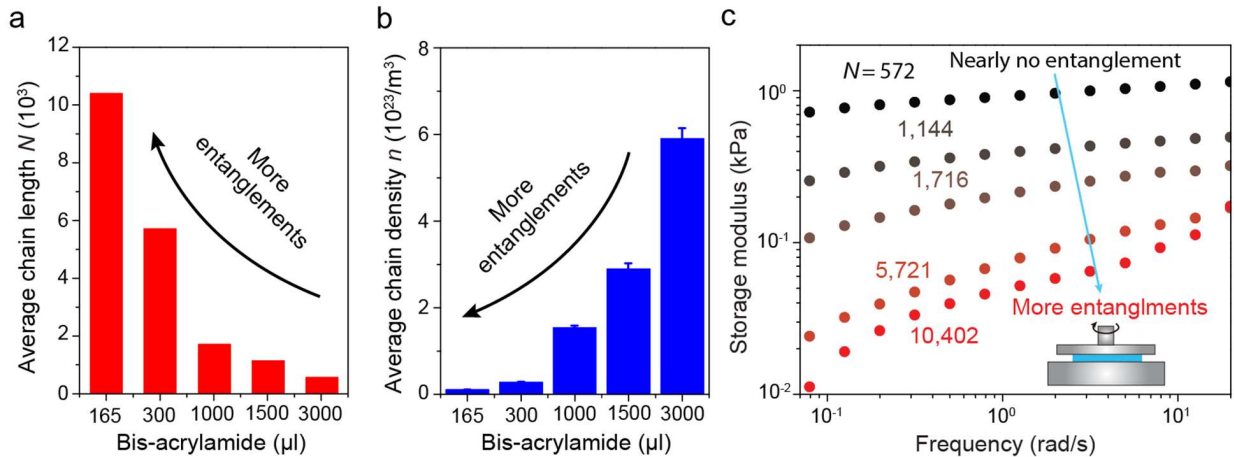

**Fig. 2. PAAm hydrogels with controlled chain entanglements.** (a) Average chain length $N$ of PAAm hydrogels with various amounts of 0.23 wt.% crosslinkers (i.e., Bis-acrylamide) in 13 ml total pre-gel solutions. (b) Average chain density $n$ of PAAm hydrogels in the swollen state with various amounts of 0.23 wt.% crosslinkers (i.e., Bis-acrylamide) in 13 ml total pre-gel solutions. (c) Storage modulus versus angular frequency of PAAm hydrogels with various chain lengths, measured in swollen state.

**Table 2. Compositions and parameters of PAAm hydrogels with controlled chain entanglements.**

| Amounts of crosslinkers* (μl) | $\phi_p$ (%) | $\phi_c$ (%) | $N$ | $NM_m$ (g/mol) | $n$ ($10^{23}/m^3$) | $n$ (mol/m³) | $\lambda_s$ |
|---|---|---|---|---|---|---|---|
| 165 | 14 | 0.0029 | 10,402 | 738,542 | 0.11 | 0.02 | 2.20 |
| 300 | 14 | 0.0053 | 5,721 | 406,191 | 0.28 | 0.05 | 1.95 |
| 1000 | 14 | 0.0177 | 1,716 | 121,836 | 1.54 | 0.26 | 1.65 |
| 1500 | 14 | 0.0265 | 1,144 | 81,224 | 2.90 | 0.48 | 1.53 |
| 3000 | 14 | 0.0531 | 572 | 40,612 | 5.91 | 0.98 | 1.52 |

* denotes the amounts of 0.23 wt.% bis-acrylamide in 13 ml pre-gel solutions.

## 2.2 Rheological characterization of chain entanglement

We next perform rheological tests to characterize the level of chain entanglement in PAAm hydrogels with different average chain lengths in the swollen state. We cut the swollen hydrogels into disk-shaped samples with diameters of 10 mm. The thicknesses of the samples are fixed at 1.5 mm in the as-prepared state. The maximum oscillatory shear stress is controlled as 5 Pa, measuring the rheological properties (e.g., storage modulus, loss modulus) at small deformations. As shown in **Fig. 2c**, PAAm hydrogel with a short average chain length (i.e., $N = 572$) shows a negligible rate dependence at the angular frequencies from 0.05 to 20 rad/s. As the average chain length increases, the rate dependence of storage modulus becomes more and more pronounced.

The rate dependence of storage modulus in PAAm hydrogels can be attributed to multiple possible physical causes, including migration of water molecules [28], dynamics of reversible bonds [29, 30], and/or reptation of entangled polymer chains [31]. Here, we can exclude the possibility of migration of water molecules and dynamics of reversible crosslinks because of the following reasons. First, since the loading mode in our rheology test is simple shear, there is no hydrostatic pressure applied on the sample to drive the migration of water molecules. Second, PAAm hydrogels are covalently crosslinked free of reversible bonds. Therefore, the rate dependence of storage modulus in PAAm hydrogels is largely attributed to the reptation of entangled polymer chains. To summarize, the controlled reduction of crosslink densities in a PAAm hydrogel can effectively produce chain entanglement by increasing its average chain length. We denote the PAAm hydrogel with the average chain length of $N = 572$ as a nearly unentangled polymer network and the PAAm hydrogel with the average chain length of $N = 10,402$ as an entangled polymer network.

## 2.3 Comparison between fracture toughness and fatigue threshold

We use the pure shear tensile tests to measure the fracture toughness of PAAm hydrogels with controlled levels of chain entanglement (**Fig. 3a**). We first measure the nominal stress $s$ versus stretch $\lambda$ curve of an unnotched sample as plotted in **Fig. 3b**. We further introduce a sharp crack in the other sample with the same dimensions as the notched sample and measure the critical stretch (i.e., $\lambda_c$), at which crack propagates steadily. The measured fracture toughness can be calculated through $\Gamma_{\text{fracture}} = H \int_1^{\lambda_c} s \, d\lambda$, where $H$ is the height of the sample. **Figure 3c** summarizes the measured fracture toughness of PAAm hydrogels with various average chain lengths. Similar to the rate-dependent fracture toughness in highly entangled elastomers [32], the fracture toughness of entangled PAAm hydrogel shows slight rate dependence as the loading rate increases from 0.5 min$^{-1}$ to 54 min$^{-1}$ (**Fig. A2**). Notably, the rate dependence of fracture toughness in entangled PAAm hydrogel is much less pronounced compared with entangled elastomers. This is possibly because the swollen PAAm hydrogels contain a large quantity of water molecules that screen the intermolecular interactions between PAAm polymer chains, thereby decreasing the rate dependence of entangled polymer networks.

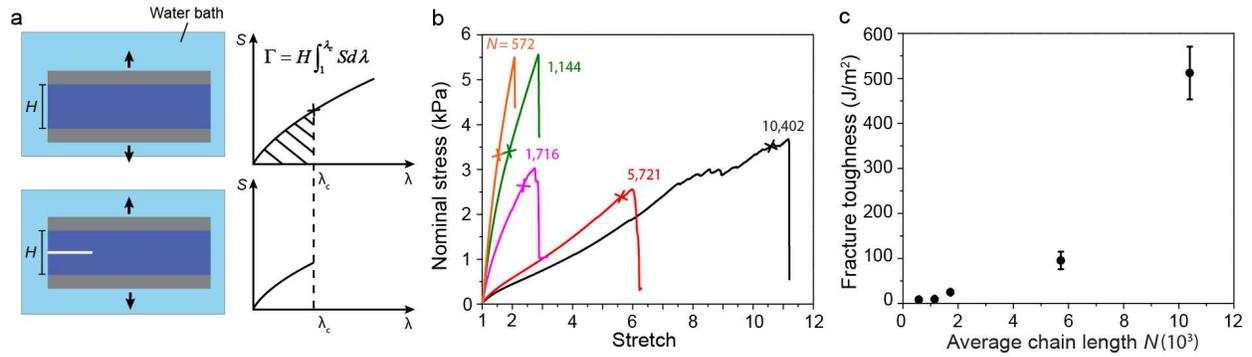

**Fig. 3. Fracture characterization of PAAm hydrogels with various average chain lengths.** (a), Schematic illustration of the pure shear tensile test to measure the fracture toughness $\Gamma_{\text{fracture}}$. (b), Nominal stress versus stretch curves of PAAm hydrogels with different average chain lengths. The cross points denote the critical stretch $\lambda_c$ at which crack propagates in the notched samples. (c), Fracture toughness $\Gamma_{\text{fracture}}$ versus average chain length $N$.

We further perform fatigue tests to measure the fatigue threshold of PAAm hydrogels with controlled levels of chain entanglement. As schematically illustrated in **Fig. 4a**, we cyclically load an unnotched sample to measure the nominal stress versus stretch curve (i.e., $s$ vs. $\lambda$) under pure shear tensile loading. The strain energy density stored in the sample over cycles can be calculated through $W(\lambda_{\text{applied}}) = \int_1^{\lambda_{\text{applied}}} s\, d\lambda$, where $\lambda_{\text{applied}}$ is the applied stretch. Similar to the fracture test, we use a razor blade to introduce a sharp crack in the sample. We perform cyclic loading on the other notched sample with the same dimensions as the unnotched sample and further use a camera (Imaging Source) to record the crack extension (i.e., $\Delta c$) over cycles (i.e., $N_{\text{cycle}}$). As the camera resolution is around 30 μm/pixel and we apply 1000 cycles of loading to measure the crack extension per cycle $dc/dN$, the lower bound of the detectable value of $dc/dN$ is around 30 nm/cycle. The applied energy release rate can be calculated through $G(\lambda_{\text{applied}}) = HW(\lambda_{\text{applied}})$, where $H$ is the gage length of the notched sample and $\lambda_{\text{applied}}$ is the applied stretch. When the applied energy release rate $G$ is small, there is no observed crack extension, giving the fatigue crack extension rate as zero (i.e., $dc/dN_{\text{cycle}} = 0$). As the applied energy release rate increases, we can observe a fatigue crack extension over cycles. The applied stretch and applied energy release rate for all samples are provided in **Table A1**. The slope of the fatigue crack extension curve versus cycle number gives a finite fatigue crack extension rate $dc/dN_{cycle}$. By linearly extrapolating the curve to the $G$ axis, we identify a critical energy release relate $G_c$ as the measured fatigue threshold of the sample (i.e., $\Gamma_{fatigue} = G_c$) as shown in **Fig. A3**. **Figure 4b** summarizes the measured fatigue threshold as a function of average chain length. The fatigue threshold increases with the average chain length, which is consistent with the classical Lake-Thomas model [4] as derived in **Appendix B**.

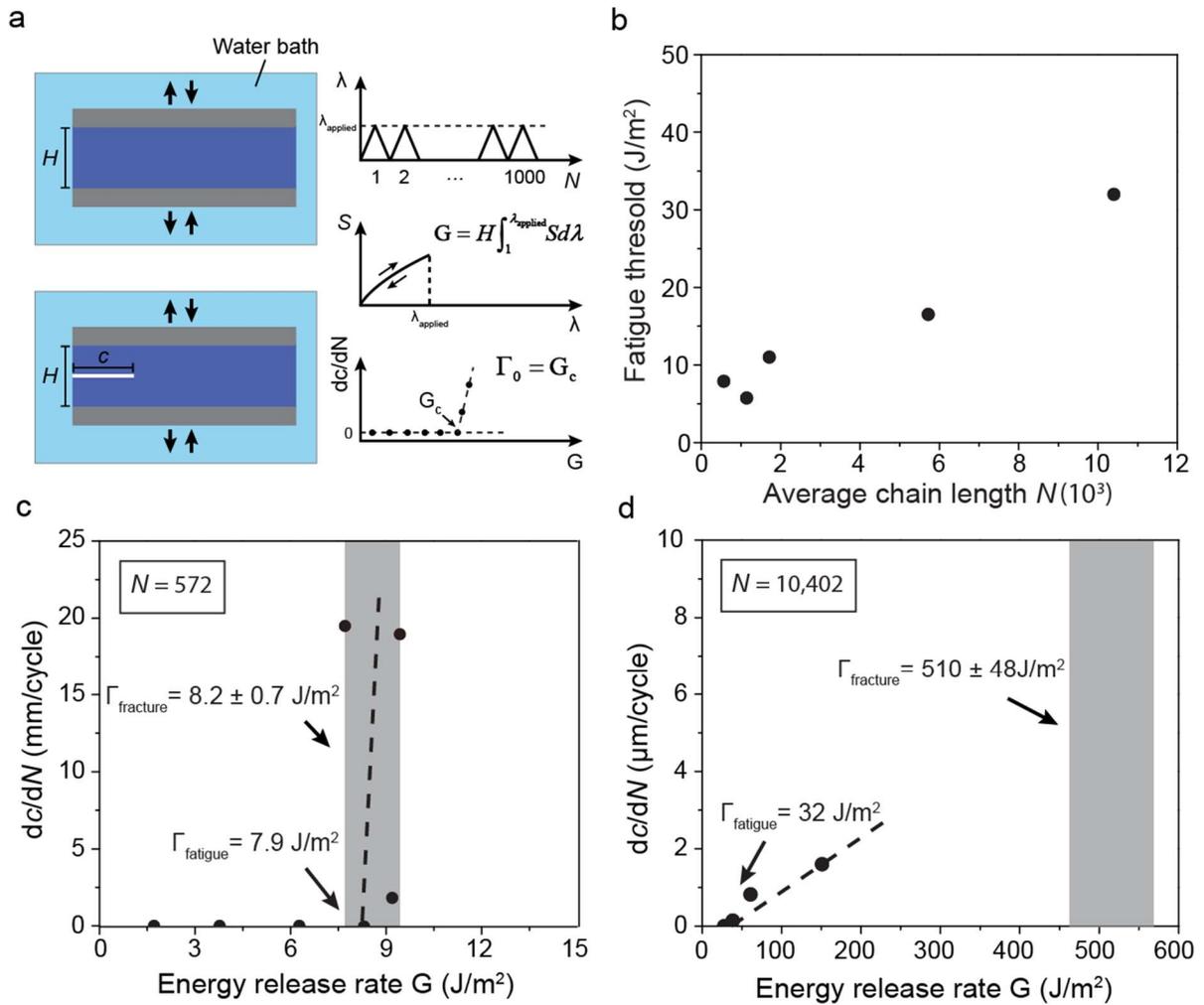

**Fig. 4. Fatigue characterization of PAAm hydrogels with various average chain lengths.** (a), Schematic illustration of the pure shear tensile test to measure the fatigue threshold. (b), Summarized fatigue threshold $\Gamma_{fatigue}$ versus average chain length $N$ of PAAm hydrogels with controlled chain entanglements. (c), Fatigue crack extension rate $dc/dN_{cycle}$ versus applied energy release rate $G$ of the nearly unentangled polymer network (i.e., PAAm hydrogel with average chain length $N = 572$). The measured fracture toughness (i.e., $\Gamma_{fracture} = 8.2 \pm 0.7 \text{ J/m}^2$) is almost the same as its fatigue threshold (i.e., $\Gamma_{fatigue} = 7.9 \text{ J/m}^2$). (d), Fatigue crack extension rate $dc/dN_{cycle}$ versus applied energy release rate $G$ of the entangled polymer network (i.e., PAAm hydrogel with average chain length $N = 10,402$). The measured fracture toughness (i.e., $\Gamma_{fracture} = 510 \pm 48 \text{ J/m}^2$) is about 16 times larger than its fatigue threshold (i.e., $\Gamma_{fatigue} = 32 \text{ J/m}^2$).

**Figure 4c** plots the fatigue crack extension curve versus the applied energy release rate of a nearly unentangled polymer network (i.e., PAAm hydrogel with the average chain length $N = 572$), measuring its fatigue threshold as 7.9 J/m². The measured fracture toughness of the same material is measured as $8.2 \pm 0.7$ J/m² as highlighted by the grey region in **Fig. 4c**, which is almost the same as its fatigue threshold. Our data indicate that a nearly unentangled polymer network has almost the same fracture toughness and fatigue threshold. Different from the ideal polymer network which also has the identical fracture toughness and fatigue threshold [27], the nearly unentangled polymer network (i.e., PAAm hydrogel with the average chain length $N = 572$) still contains other forms of non-ideal features including non-uniform functionality, non-uniform chain length, dangling chains, and cyclic loops. This observation indicates that the non-ideal features in an unentangled polymer network do not account for the difference between fracture toughness and fatigue threshold. **Figure 4d** plots the fatigue crack extension curve versus the applied energy release rate of an entangled polymer network (i.e., PAAm hydrogel with the average chain length $N = 10,402$). Its fatigue threshold is measured as 32 J/m², slightly larger than that of the nearly unentangled polymer network (i.e., 7.9 J/m²), but 16 times lower than its fracture toughness (i.e., $510 \pm 48$ J/m²) as highlighted by the grey region in **Fig. 4d**. The presence of chain entanglements in PAAm hydrogels results in a huge difference between fracture toughness and fatigue threshold.

## 2.4 Characterization of the stress-stretch hysteresis

We further perform cyclic loading-unloading tests to measure the stress-stretch hysteresis ratio of both entangled and nearly unentangled polymer networks. The maximum hysteresis ratio $h_{max}$ is defined as $h_{max} = U_D / U_{max}$, where $U_D = \oint_1^{\lambda_{max}} s d\lambda$ is the maximum mechanical dissipation per unit volume of the material and $U_{max} = \int_1^{\lambda_{max}} s d\lambda$ the maximum mechanical work done on unit volume of the material with $s$ being the nominal stress, $\lambda$ being the stretch, $\lambda_{max}$ being the maximum stretch at which the sample fails. As shown in **Fig. 5a**, the presence of chain entanglement slightly increases the maximum hysteresis ratio from 5% to 10%. We further compare the stress versus stretch curves under cyclic loading at various loading rates. For the nearly unentangled polymer network, its stress versus stretch curve is nearly independent of the loading rate (**Fig. 5b**). For the entangled polymer network (**Fig. 5c**), the modulus from the stress versus stretch curve shows a rate dependence, which is consistent with the rheology characterization. Intriguingly, the maximum hysteresis ratios of the entangled polymer network at different loading rates (i.e., 0.06, 0.12, 0.25, 0.46 s$^{-1}$) are almost the same (**Fig. B1**).

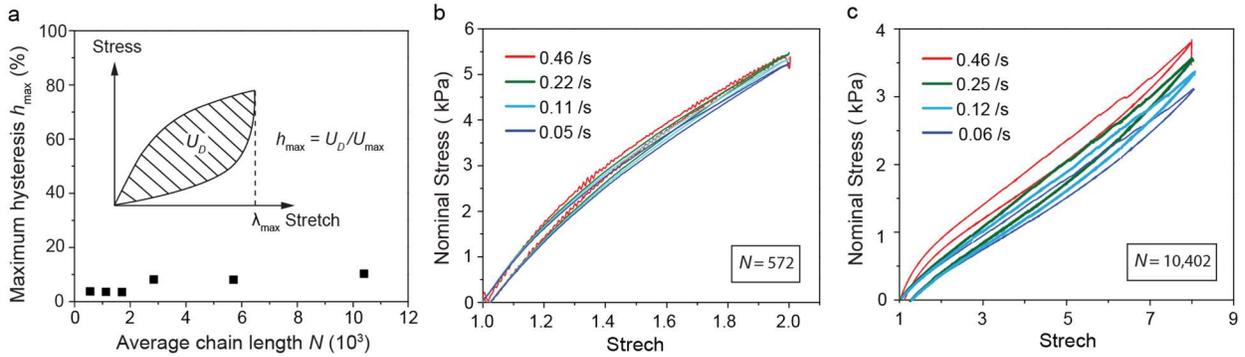

**Fig. 5. Characterization of the stress-stretch hysteresis ratio in entangled and unentangled polymer networks.** (a), The maximum hysteresis ratio $h_{max}$ versus average chain length $N$. (b) Nominal stress versus stretch curve of PAAm hydrogels with the average chain length of 572 (representing a nearly unentangled polymer network) at various loading rates of 0.05, 0.11, 0.22, and 0.46 s$^{-1}$. (c) Nominal stress versus stretch curve of PAAm hydrogels with the average chain length of 10,402 (representing an entangled polymer network) at various loading rates of 0.06, 0.12, 0.25, and 0.46 s$^{-1}$.

## 2.4 Bulk dissipation model

To understand the near-crack toughening mechanism in an entangled polymer network, we first review the bulk dissipation model [15, 16]. The bulk dissipation model describes how a large stress-stretch hysteresis of a bulk material toughens the material. Once a crack propagates in a soft tough material, there are two physical processes. First, the scission of a layer of polymer chains on the crack path provides the intrinsic fracture energy of the material $\Gamma_0$, following the Lake-Thomas model [4]. Physically, the intrinsic fracture energy of a soft material is identical to its fatigue threshold (i.e., $\Gamma_0 = \Gamma_{fatigue}$) [5]. Second, material elements in a process zone around the crack will experience loading and unloading as the crack propagates, which dissipating substantial mechanical energy. We denote the contribution of the bulk hysteretic mechanical dissipation to the fracture toughness as $\Gamma_D^{bulk}$. Therefore, the total fracture toughness of a soft material can be expressed as

$$\Gamma_{fracture} = \Gamma_0 + \Gamma_D^{bulk} \tag{1}$$

where $\Gamma_D^{bulk} = U_D l_D$ with $U_D$ being the mechanical energy dissipated per the volume of the process zone, and $l_D$ being the size of the process zone. $U_D$ is a measurable quantity defined as $U_D = \oint_1^{\lambda_{max}} s d\lambda$, where $\lambda_{max}$ is the maximum stretch at which the sample fails.

To have the explicit expression of the fracture toughness of a soft material, one needs the stress distribution profile around the crack tip to estimate the size of the process zone. We take the soft material as a neo-Hookean solid, giving the leading order of the nominal stress at a point near the crack tip scales as $s \propto \sqrt{\Gamma_{fracture} \mu / x}$, where $\mu$ is the shear modulus of the materials and $x$ is the distance from the point to the crack tip [33]. We choose the maximum stress $s_{max}$ to determine the boundary of the process zone, therefore the size of the process zone scales as

$$l_D \propto \Gamma_{fracture} \mu / s_{max}^2 \propto \Gamma_{fracture} / U_{max} \tag{2}$$

where $U_{max} \propto S_{max}^2/\mu$ is the maximum mechanical work done on the material. The size of the process zone for soft tough materials is typically greater than 100 μm [34, 35]. A combination of Eqs. (1) and (2) gives the explicit expression for the toughness enhancement of a soft material as

$$\frac{\Gamma_{fracture}}{\Gamma_0} = \frac{1}{1-\alpha h_{max}} \quad (3)$$

where $h_{max} = U_D/U_{max}$ is the maximum hysteresis defined as the ratio between the maximum dissipation and the maximum mechanical work done on the material, $0 \leq \alpha \leq 1$ is a dimensionless parameter depending on the stress-stretch hysteresis of the material deformed to different levels of stretch ($\alpha = 1$ for highly stretchable materials such as the PAAm hydrogels).

Given the measured $h_{max}$, we can use the bulk dissipation model (i.e., Eq. (3)) to calculate the toughness enhancement, which is consistently below 1.1 for PAAm hydrogels with various levels of chain entanglement (**Fig. 6b**). In contrast, the measured toughness enhancement for PAAm hydrogels with high average chain length (i.e., entangled polymer network) is as large as 16. The discrepancy between the experimental result and bulk dissipation model's prediction indicates that the bulk dissipation model fails to explain the low-hysteresis toughness enhancement in entangled polymer networks. We attribute the low-hysteresis toughness enhancement of soft materials to a new mechanism, *near-crack dissipation*, as discussed in the next section.

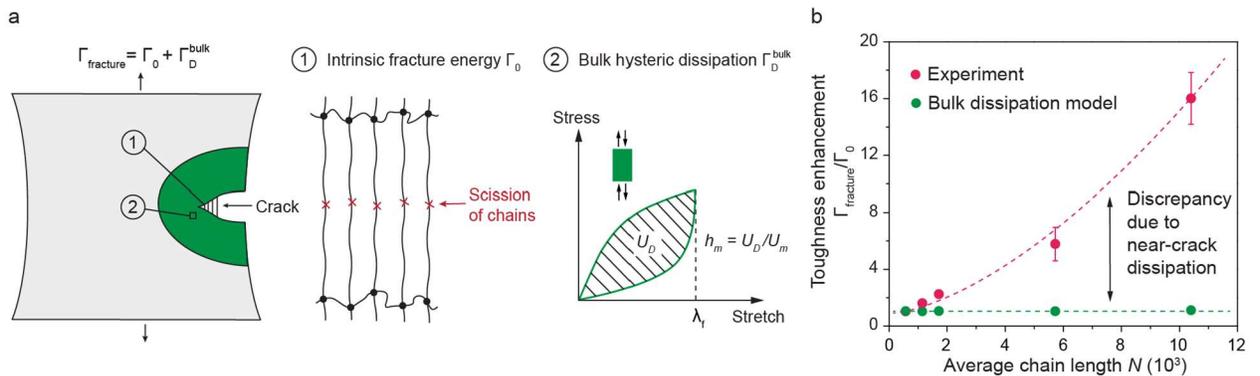

**Fig. 6. Discrepancy of toughness enhancement between the bulk dissipation model and experimental results.** (a) The bulk dissipation model relies on large stress-stretch hysteresis of the bulk material. The size of the process zone in the bulk dissipation model is typically larger than hundreds of micrometers. (b) Comparisons between the experimentally measured toughness enhancement and the predicted toughness enhancement by the bulk dissipation model as a function of average chain length $N$.

**2.5 Chain entanglement gives near-crack dissipation**

In this paper, we propose a new toughening mechanism, *near-crack dissipation*, to account for the low-hysteresis toughness enhancement in entangled polymer networks (**Fig. 7**). Once a crack propagates in an entangled polymer network, the highly entangled polymer chains across the crack plane are pulled out, potentially dissipating substantial energy due to abundant intermolecular interactions between neighboring chains. In addition, once the entangled chains around the crack tip are highly stretched, scissions of chains can be delocalized to multiple adjacent layers around crack plane, dissipating more energy than fracturing a single layer of chains (**Fig. 7** and **Fig. A4**) [32]. Notably, the stretch applied on the bulk entangled polymer network before failure can be much lower than the stretch of the crack tip. Therefore, the pull-out and delocalized damage of chains in the bulk entangled polymer network under stretches may be negligible, leading to low stress-stretch hysteresis of the bulk network (**Fig. 7**). Overall, the fracture toughness $\Gamma_{fracture}$ of an entangled yet low-hysteresis polymer network is equal to the summation of its intrinsic fracture energy $\Gamma_0$ and its dissipative fracture energy due to pull-out and delocalized damage of chains near the crack tip $\Gamma_D^{tip}$,

$$\Gamma_{fracture} = \Gamma_0 + \Gamma_D^{tip} \quad (4)$$

Recent experiment indeed observed the delocalized scission of polymer chains around crack path using mechanophores in an entangled elastomer (**Fig. A4**) [32]. The measurement reveals that bond scission, far from being restricted to a constant level near the crack plane, can be delocalized over up to hundreds of micrometers (**Fig. 7c**). To further gain molecular insights on the near-crack dissipation process, future efforts will be focused on experimental characterization of such pull-out of entangled chains and delocalized damage of chains at the crack tip of entangled polymer networks.

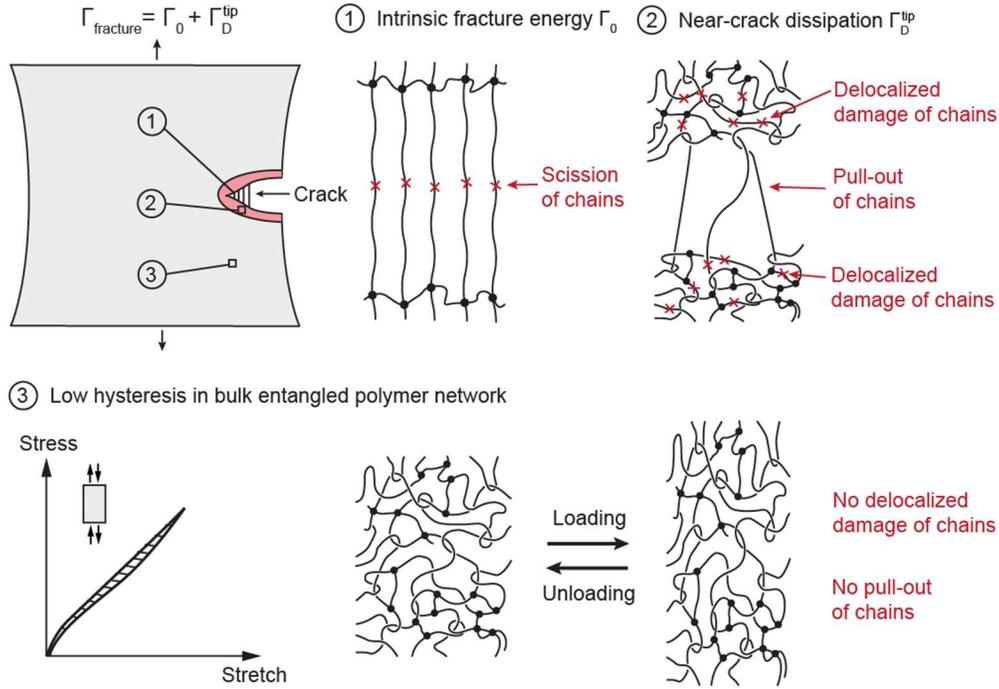

**Fig. 7. Schematic illustration of the near-crack dissipation mechanism for entangled polymer network.** Pull-out and delocalized damage of chains around the crack tip can dissipate substantial energy, toughening the entangled polymer network. The stretch applied on the bulk entangled polymer network before failure can be much lower than the stretch of the crack tip, giving negligible pull-out and delocalized damage of chains and thus low stress-stretch hysteresis of the bulk network.

## 3. Conclusion Remarks

In this work, we use polyacrylamide hydrogels as a model material to systematically investigate fracture and fatigue in entangled and unentangled polymer networks. We find that the fracture toughness and the fatigue threshold of a nearly unentangled polymer network are almost the same (i.e., $\Gamma_{fracture} = \Gamma_{fatigue}$), although the polymer network still contains non-ideal features including structural heterogeneity (i.e., non-uniform chain lengths, non-uniform functionalities) and topological defects (i.e., dangling chains, and cyclic loops). In contrast, for an entangled polymer network, its fracture toughness is 16 times larger than its fatigue threshold (i.e., $\Gamma_{fracture} > \Gamma_{fatigue}$), indicating a significant toughness enhancement due to chain entanglement. More intriguingly, the toughness enhancement in a highly entangled polymer network requires low stress-stretch hysteresis of the bulk network (<10%), which is contradictory to the well-known bulk dissipation model. We attribute the low-hysteresis toughness enhancement in entangled polymer networks to a new toughening mechanism, *near-crack dissipation*, possibly induced by pull-out of chains and delocalized damage of chains around the crack tip. This work not only reveals the effect of chain entanglement on fracture and fatigue of polymer networks but also suggests routes for the design of low-hysteresis soft yet tough materials [20, 36].

**Appendix A. Supporting experimental data**

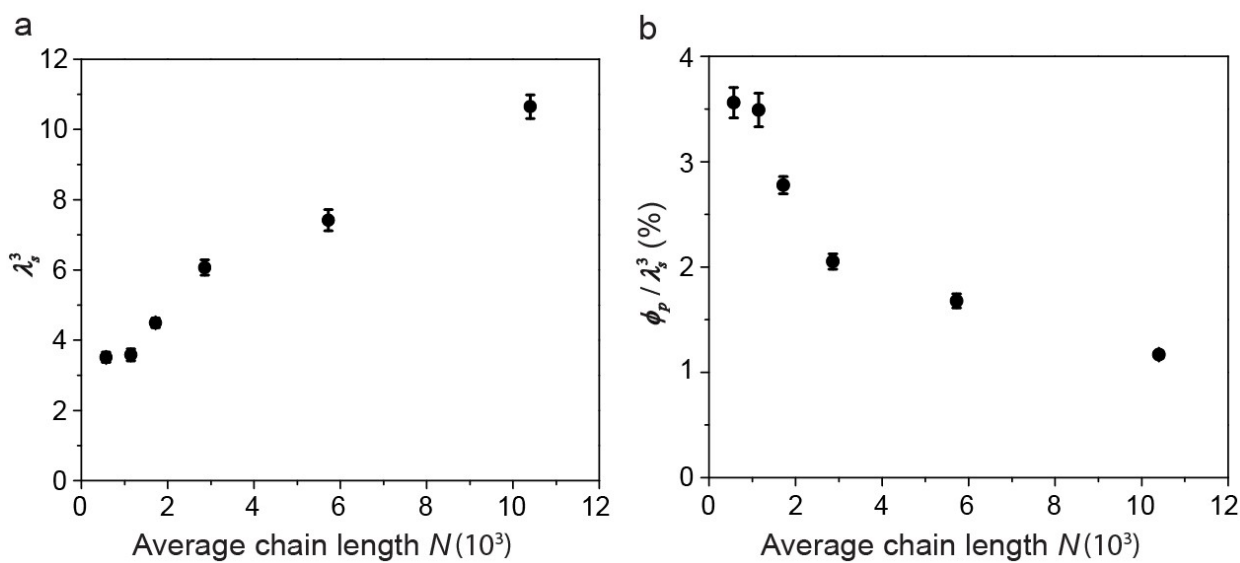

**Fig. A1. Summarized parameters for PAAm hydrogels with various average chain lengths.** (a), Swelling ratio in volume $\lambda_s^3$ versus average chain length $N$. (b), Polymer concentration in the swollen state $\phi_p / \lambda_s^3$ versus average chain length $N$.

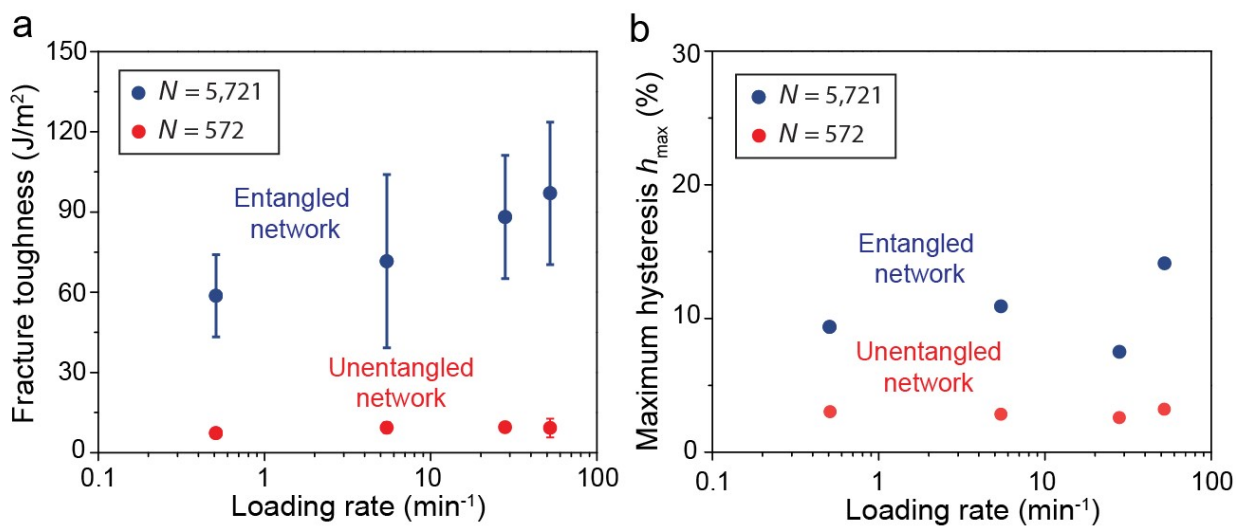

**Fig. A2. Characterization the rate effects on fracture toughness and hysteresis.** (a), Fracture toughness $\Gamma_{fracture}$ versus loading rate of a nearly unentangled polymer network (i.e., $N = 572$) and an entangled polymer network (i.e., $N = 5,721$). (b), Maximum hysteresis ratio $h_{max}$ versus loading

rate of a nearly unentangled polymer network (i.e., $N = 572$) and an entangled polymer network (i.e., $N = 5,721$).

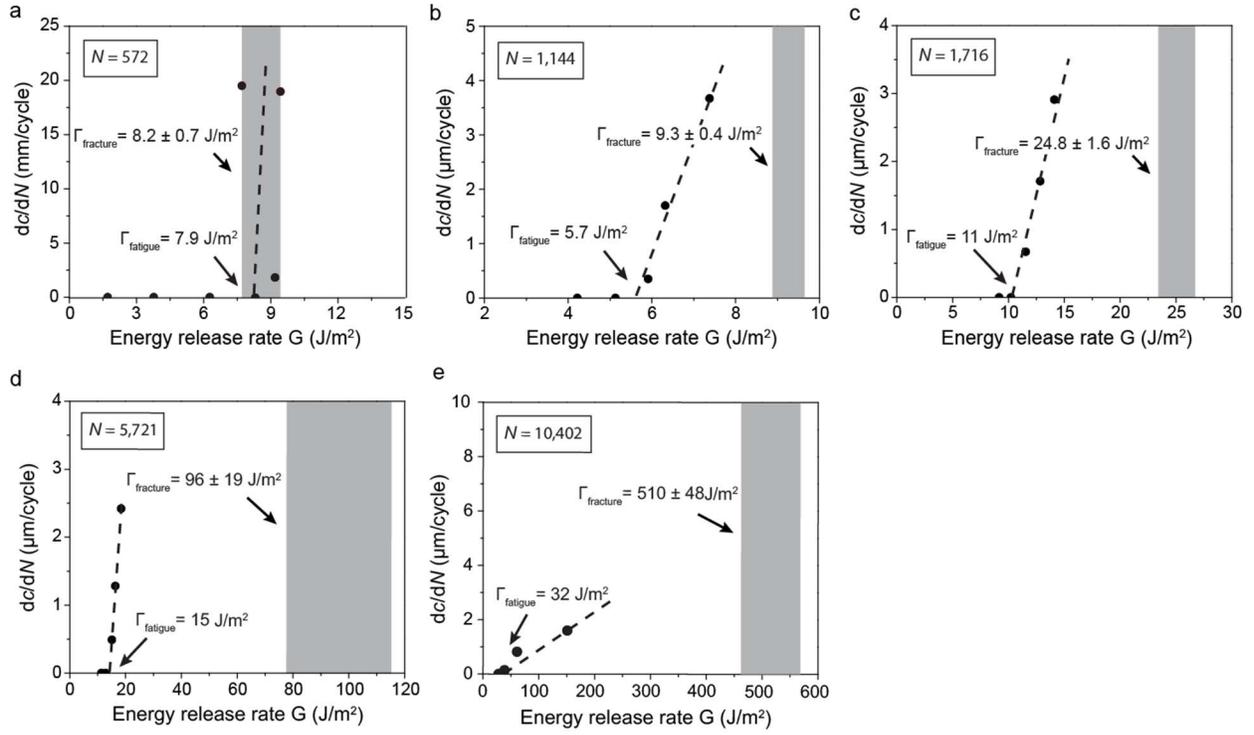

**Fig. A3. Summarized fatigue-induced crack extension versus applied energy release rate for hydrogels with various chain entanglements.** (a) $N = 572$. (b) $N = 1,144$. (c) $N = 1,716$. (d) $N = 5,721$. (e) $N = 10,402$.

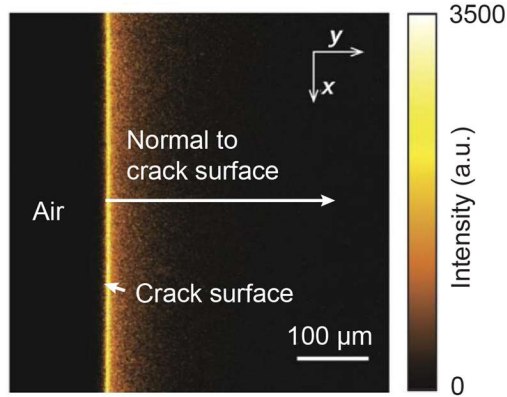

**Fig. A4.** Damage quantification through confocal imaging using mechanophores in an entangled elastomer [32]. The measurement reveals that bond scission, far from being restricted to a constant level near the crack plane, can be delocalized over up to hundreds of micrometers. Image is reprinted with permission from *Slootman et al., 2020, Quantifying Rate- and Temperature-Dependent Molecular Damage in Elastomer Fracture, Phys. Rev. X 10: 041045*.

**Table A1. Applied stretch and applied energy release rate in the fatigue tests.**

| | | | | | |
|---|---|---|---|---|---|
| $N = 572$ | | | | | |
| $\lambda_{applied}$ | 1.37 | 1.51 | 1.54 | 1.56 | 1.63 |
| $G$ (J/m$^2$) | 3.77 | 6.42 | 7.33 | 7.89 | 9.61 |
| $N = 1{,}144$ | | | | | |
| $\lambda_{applied}$ | 1.48 | 1.53 | 1.58 | 1.60 | 1.66 |
| $G$ (J/m$^2$) | 4.22 | 5.13 | 5.91 | 6.32 | 7.38 |
| $N = 1{,}716$ | | | | | |
| $\lambda_{applied}$ | 1.73 | 1.79 | 1.85 | 1.90 | 1.95 |
| $G$ (J/m$^2$) | 9.17 | 10.22 | 11.55 | 12.84 | 14.11 |
| $N = 5{,}721$ | | | | | |
| $\lambda_{applied}$ | 2.20 | 2.29 | 2.42 | 2.49 | 2.61 |
| $G$ (J/m$^2$) | 11.35 | 12.88 | 15.11 | 16.35 | 18.44 |
| $N = 10{,}402$ | | | | | |
| $\lambda_{applied}$ | 2.84 | 3.27 | 3.54 | 4.36 | 6.31 |
| $G$ (J/m$^2$) | 24.33 | 35.24 | 42.71 | 69.34 | 152.07 |

**Appendix B. Theory for fatigue thresholds**

Using recently developed defect-network fracture model [37], we can estimate the fatigue threshold of PAAm hydrogels at the as-prepared state as

$$\Gamma_0 = \beta n_{el} N^{3/2} b U \tag{B1}$$

where $\beta$ is a dimensionless parameter that accounts for the network architecture's contribution to the fatigue threshold [37], $N$ is the average chain length (i.e., number of monomers between neighboring crosslinkers) of PAAm polymer chains, $n_{el}$ is the average number of elastically active chains per unit volume in the as-prepared state, $b$ is the length of one AAm monomer, $U$ is the bond dissociation energy of one AAm monomer at fracture.

We next estimate the values of parameters in Eq. (B1). Given the mass fraction of polymers at the as-prepared state $\phi_p$ and the average chain length $N$, we can calculate the number of polymer chains per unit volume in the as-prepared state as $\rho_g \phi_p N_A / (N M_m)$, where $\rho_g = 10^3 \, \text{kg/m}^3$, $\phi_p = 0.14$, $M_m = 71 \, \text{g/mol}$ is the molecular weight of the monomer, and $N_A = 6.02 \times 10^{23} \, \text{mol}^{-1}$ is Avogadro number. Presuming there are negligible inactive chains in PAAm hydrogels, we can regard all the polymer chains as elastically active polymer chains, namely $n_{el} = \rho_g \phi_p N_A / (N M_m)$. The length of one AAm monomer is estimated as $b = 0.434 \, \text{nm}$ [22]. Since the backbone of one AAm monomer contains one C-C bond, the bond dissociation energy of one AAm monomer is estimated as the bond energy of C-C bond (i.e., 346 kJ/mol), giving $U = 346 \, \text{kJ/mol}$. By substituting the values and expressions of $n$, $b$, and $U$, we can derive the expression of fatigue threshold of PAAm hydrogels in the as-prepared state as

$$\Gamma_0 = 0.3 \beta N^{1/2} \tag{B2}$$

As shown in **Fig. B1**, by fitting the experimental data with our theory in Eq. (B2), we can identify the dimensionless parameter $\beta = 3.9$.

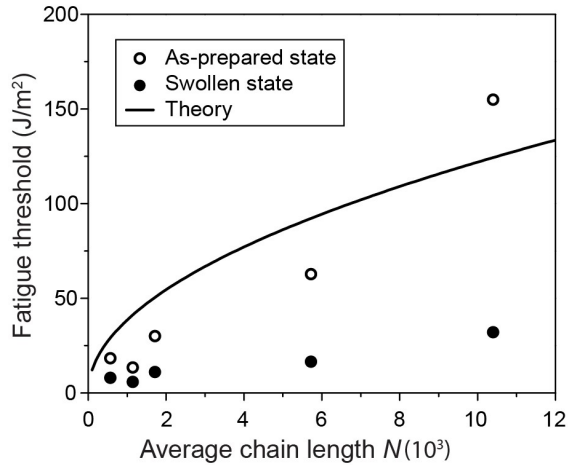

**Fig. B1. Fatigue thresholds for PAAm hydrogels with various average chain lengths.** The solid dots denote the fatigue threshold of PAAm hydrogels in the swollen state. The hollow dots denote the fatigue threshold of PAAm hydrogels in the as-prepared state. The solid line denotes the theoretical calculation of fatigue threshold of PAAm hydrogels in the as-prepared state.

## Acknowledgements

This work is supported by the U.S. Army Research Office through the Institute for Soldier Nanotechnologies at MIT (W911NF-13-D-0001).

## Declaration of competing interest

The authors declare that they have no known competing financial interests or personal relationships that could have appeared to influence the work reported in this paper.

## Credit authorship contribution statement

Dongchang Zheng and Shaoting Lin contributed equally to this work. Dongchang Zheng and Jiahua Ni synthesized the samples. Dongchang Zheng conducted the fracture and fatigue tests Shaoting Lin and Dongchang Zheng analyzed and processed the data. Shaoting Lin drafted the paper with comments from all authors. Xuanhe Zhao supervised the study.